\documentclass[11pt]{article}

\usepackage[top = 30 mm, bottom = 30mm, left = 20mm, right=20mm]{geometry}
\usepackage{amsmath, amssymb}
\usepackage[english]{babel}
\usepackage[utf8]{inputenc}
\usepackage{latexsym}
\usepackage{graphicx}
\usepackage{lmodern}
\usepackage{jheppub}
\usepackage{slashed}
\usepackage{charter}
\usepackage{xcolor}
\usepackage{color}
\usepackage{float}
\usepackage{bm}

\numberwithin{equation}{section}

\newcommand{\be}{\begin{equation}}
\newcommand{\ee}{\end{equation}}
\newcommand{\bear}{\begin{array}}
\newcommand{\eear}{\end{array}}

\title{Late time acceleration in Palatini gravity}
\author[a, b, c]{Ignatios Antoniadis}
\author[a]{Anthony Guillen}
\author[d, e]{Kyriakos Tamvakis}

\affiliation[a]{Laboratoire de Physique Th\'eorique et Hautes Energies - LPTHE\\
Sorbonne Universit\'e, CNRS, 4 Place Jussieu, 75005 Paris, France}
\affiliation[b]{SISSA, Via Bonomea 265, 34136, Trieste, Italy}
\affiliation[c]{ICTP, Strada Costiere 11, 34151 Trieste, Italy}
\affiliation[d]{Physics Department, University of Ioannina, 45110, Ioannina, Greece}
\affiliation[e]{CERN,Theoretical Physics Department, 1211 Geneva 23, Switzerland}

\emailAdd{antoniad@lpthe.jussieu.fr}
\emailAdd{anthony.guillen@protonmail.com}
\emailAdd{tamvakis@uoi.gr}

\date{} 

\abstract{We investigate the effect of the quadratic correction $\alpha R^2$ and non-minimal coupling $\xi \phi^2 R$ on a quintessence model with an exponential potential $V(\phi) = M^4\exp(-\lambda\phi)$ in the Palatini formulation of gravity. We use dynamical system techniques to analyze the attractor structure of the model and uncover the possible trajectories of the system. We find that the quadratic correction cannot play a role in the late time dynamics, except for unacceptably large values of the parameter $\alpha$; although it can play a role at early times. We find viable evolutions, from a matter-dominated phase to an accelerated expansion phase, {with the dynamics} driven by the non-minimal coupling. {These evolutions correspond} to trajectories where the field ends up frozen, thus acting as a cosmological constant.}

\begin{document}

\maketitle\newpage

\section{Introduction}

\vfill

Our understanding of the primordial Universe relies heavily on the inflationary paradigm \cite{AAS, DK, KS, AHG, ADL1, AS, ADL2}, according to which it went through a phase of exponential expansion. This circumvents fine-tuning issues of the hot big bang model and provides a mechanism for the generation of the cosmological perturbations responsible for the large scale structure of the Universe \cite{AAS1, MUK, HAWK, AAS2, AHG2, BST}. Strong evidence in favour of primordial inflation is supplied by ongoing observations of increasing accuracy \cite{PLANCK, BICEP-WMAP}. Most simple models of inflation feature a scalar field, named the {\textit{inflaton}}, whose potential energy dominates the Universe during the inflationary phase of exponential expansion. 

\vfill

However, observations suggest that there is a second phase of accelerated expansion taking place at present and arising from an unknown source of energy, called {\textit{dark energy}}, which makes up $70\%$ of the present energy content of the Universe. These observations are consistent with dark energy being a positive cosmological constant, but one has to explain the vanishingly small value of this constant in natural units. A simple alternative is to consider that the late time accelerated expansion also comes from the dynamics of a scalar field, named {\textit{quintessence}} \cite{RP, CDS, CST, Ferreira:1997hj}. Quintessence models introduce new issues related to initial conditions. Basically one has to explain the transition from matter domination to the accelerated expansion era. Such a transition can be naturally obtained if the model is endowed with the right attractor structure. 

\vfill

A rather more ambitious set of models, called models of {\textit{quintessential inflation}}, unify both phases of accelerated expansion, identifying the quintessence field with the inflaton itself \cite{PV}. In these models the inflaton has survived down to the present epoch, {and the process of reheating is attributed} to extra fields and mechanisms. Therefore, considerable uncertainties are associated with such an ambitious scheme. 

\vfill

The common ground of all cosmological models is Einstein's theory of General Relativity, based of the Einstein-Hilbert action, with matter fields minimally coupled to the metric. Nevertheless, quantum interactions of these fields do generate corrections to the action, such as a quadratic correction in the Ricci scalar $\alpha R^2$ or non-minimal couplings of scalar fields to the Ricci scalar $\xi\phi^2R$. As a matter of fact, the former term unlocks a new gravitational scalar degree of freedom that plays the role of the inflaton in the successful Starobinsky model \cite{AAS}. The non-minimal coupling, on the other hand is at the center of Higgs inflation models \cite{BS}. It is understood that these terms, even if not included {in the classical action}, are bound to arise at the {quantum level, so they must be included in the effective gravitational theory} employed for cosmological considerations.

\vfill\newpage\vfill

In the case of pure Einstein-Hilbert gravity with minimal couplings, the standard metric formulation (in which the affine connection is constrained to take the Levi-Civita form) and the Palatini or first order formulation \cite{PALA} (in which the affine connection is treated as an independent variable and takes the form dictated by its own equation of motion) are equivalent. However, in the presence of higher order corrections or non-minimal couplings, they cease to be, leading to different predictions \cite{BD}. For example, in the Palatini version of the Starobinsky model, the term $\alpha R^2$ does not introduce a new scalar degree of freedom, contrary to its metric counterpart. Models of early inflation based on an effective theory that includes the above terms in the Palatini formulation, have been studied, with predictions compatible with observational data \cite{AKLT, AKLPT, EERW}. Quintessential inflation has also been considered in this framework \cite{Dimopoulos:2020pas, D1, D2, V}.
	
\vfill

In the present article we investigate the effect of the quadratic correction and non-minimal coupling on a simple quintessence model with an exponential potential, in the Palatini formulation of gravity. We use dynamical system techniques, reviewed in \cite{BBCC, CST}, to analyze the attractor structure of the model and uncover the possible trajectories of the system.

\vfill

We find that the quadratic correction $\alpha R^2$ does not play any role in the late-time dynamics of the system, unless the parameter $\alpha$ were to take unacceptably large values; but it could play a role at early times. However, the non-minimal coupling $\xi\phi^2R$ can drive the evolution of the system from the vicinity of an early-time matter-dominated phase towards a late-time attractor {corresponding to} accelerated expansion. {To be precise, it induces} a new local minimum of the potential, and we find that for a {\textit{``realistic"}} evolution of the Universe the only viable trajectories are the ones where the canonical field ends up {in this new minimum and acts as} a cosmological constant.

\vfill

Our paper is organized as follows: In section 2, we set up the general framework of the Palatini formulation and derive the Einstein frame equations of motion of the system for a FLRW backgound, in the presence of matter described as perfect fluid. In section 3 we focus on the quadratic correction $\alpha R^2$. We introduce a set of dynamical variables that describe the evolution of the system in terms of {a set of} first order differential equations. We analyze the fixed point structure of this system for various approximations and draw conclusions about the importance of the quadratic correction to give a {\textit{``realistic"}} evolution of the Universe. In section 4 we focus on the non-minimal coupling $\xi\phi^2R$. We express the model in terms of a canonical scalar field and study the extrema of its potential. Then, we set up the description of the system in terms of dynamical variables and study its fixed point structure. {We find realistic evolutions where the canonical field starts some distance away from the minimum of its effective potential and ends up trapped into it, acting as a cosmological constant. Linking the initial position of the field with inflation yields an interesting relation between the number of e-folds of radiation/matter domination and the energy scales of inflation/quintessence}. In section 5 we consider the {combination of both} correction terms and discuss their effect on the evolution of the Universe. Finally, in section 6 we state our conclusions.

\vfill\newpage

\section{Framework}\label{sec:framework}

We consider the action of a scalar field $\phi$, {with a self-interaction potential $V(\phi)$ and coupled to a gravity sector} that includes, next to the Einstein-Hilbert term, a quadratic Ricci scalar term $\alpha R^2$ and a non minimal coupling of the form $\xi\phi^2R$; namely
\be\label{starting_point_action}
{\cal{S}} = \int d^4x\sqrt{-g}\left\{
\frac{1}{2}(1 + \xi\phi^2)R + \frac{\alpha}{4}R^2 - \frac{1}{2}(\nabla\phi)^2 - V(\phi)\right\}.
\ee
These terms are generally expected to be generated by quantum corrections. They are also the most simple modifications that make the metric and Palatini formulations of gravity inequivalent. Similar actions have been considered for modeling primordial inflation in the metric or Palatini formulations. Our goal is to understand the influence of the nonminimal coupling and quadratic correction on late-time inflation in the framework of the Palatini formulation from the perspective of dynamical systems. Our scalar -quintessence- field is thus a generic one, that would start evolving from a nonzero vacuum expectation value after inflation and reheating. Note that we do not identify this field with the one responsible for primordial inflation ({\textit{inflaton}}), as in the so-called {\textit{quintessential inflation}} models. In what follows we shall use the mostly $+$ metric convention. Having also set $\kappa = 1$, {we will restore it} in terms of conventional units when necessary. A standard approach to tackle the $\alpha R^2$ term is to introduce an auxiliary scalar $\chi$ and rewrite the action in the equivalent form
\be{\cal{S}} = \int d^4x \sqrt{-g}\left\{
\frac{1}{2}\Omega^2R - \frac{\alpha}{4}\chi^2 - \frac{1}{2}(\nabla\phi)^2-V(\phi)\right\},
\hspace{\baselineskip}\text{where}\hspace{\baselineskip}\Omega^2 = 1+\xi\phi^2+\alpha\chi.\ee

Next, we perform a {\textit{Weyl rescaling}} of the metric to remove the non minimal coupling
\begin{equation}\label{conformal_transformation}
g_{\mu\nu} \rightarrow \Omega^{-2}g_{\mu\nu}, \hspace{\baselineskip} g^{\mu\nu} \rightarrow \Omega^{2}g^{\mu\nu}, \hspace{\baselineskip} \sqrt{-g}\rightarrow \Omega^{-4}\sqrt{-g}, \hspace{\baselineskip}R \rightarrow \Omega^2 R - \underline{3/2\Omega^{-2}(\partial\Omega^2)^2}.
\end{equation}

By doing so, we end up in the {\textit{Einstein frame}} where the fields are minimally coupled to gravity.
The underlined term in ({\ref{conformal_transformation}}), present in the metric formulation, leads to a kinetic term for $\chi$ in the Einstein frame, promoting {it} into a dynamical degree of freedom instead of an auxiliary scalar, as it was initially introduced. However, this term is absent in the Palatini formulation, and thus, $\chi$ remains auxiliary in the Einstein frame and can be integrated out algebraically as follows. Starting from the action in the Einstein frame
\be {\cal{S}_\mathrm{EH}} + {\cal{S}_\phi} = \int d^4x\sqrt{-\bar{g}}\left\{
\frac{1}{2}R - \frac{1}{2}\frac{(\nabla\phi)^2}{(1+\xi\phi^2+\alpha\chi)}-\frac{V(\phi)+\alpha/4\chi^2}{(1+\xi\phi^2+\alpha\chi)^2}\right\}, \ee

we vary with respect to $\chi$ and obtain
\be\chi = \frac{4V(\phi)+(1+\xi\phi^2)(\nabla\phi)^2}{1+\xi\phi^2-\alpha(\nabla\phi)^2}.\ee 

\vfill\newpage\vfill

Substituting this solution for $\chi$ back into the action, we get its final form as
\be\label{action_scalar_field}
{\cal{S}_\phi} = \int d^4x\sqrt{-g}\left\{
-\frac{1}{(1+\xi\phi^2)^2+4\alpha V(\phi)}\left(\frac{1}{2}(1+\xi\phi^2)(\nabla\phi)^2 + V(\phi) - \frac{\alpha}{4}(\nabla\phi)^4\right)\right\}.
\ee

\vfill

Since we want to describe a realistic Universe, we have to add to the action the contribution of (ordinary + dark) matter $\mathcal{S}_m$ (and the contribution of radiation $\mathcal{S}_r$). This is commonly described as a background perfect fluid of energy density $\rho$ and pressure density $p$, with a barotropic equation of state $\rho = w p$, where $w = 0$ ($w = 1/3$ for radiation). If we were to add  $\mathcal{S}_m$ {in} the Jordan frame, a coupling would arise between the matter energy density and the field $\phi$ in the Einstein frame, as a result of the Weyl rescaling. Such a direct coupling would introduce $\phi$-dependent modifications {to} the effective barotropic equation of state {of} matter and {affect} the duration of the matter dominated era, {in a way that is} difficult to reconcile with observations \cite{D2}. {It could also potentially lead to} {\textit{fifth-force}} type of issues. 

\vfill

{Indeed, the quintessence field generically has a mass $m_\phi\sim H_0\sim10^{-33}\ \mathrm{eV}$ (in our case for instance in Section \ref{sec:non-minimal_coupling}, we have $m_\phi\sim 10^{-33}\sqrt\xi\ \mathrm{eV}$ at present time). So this field is essentially massless, and its couplings to ordinary matter should be more than Planck suppressed to remain unobserved so far \cite{Carroll:1998zi}. In consequence, such couplings are often ignored. Yet, coupled quintessence has been studied in \cite{Amendola:1999er}, with a coupling at the level of the energy-momentum tensor of the form $\partial^\mu T_{\mu\nu, (\phi)} = C T_{(m)}\partial_\nu \phi$ and $\partial^\mu T_{\mu\nu, (m)} = -C T_{(m)}\partial_\nu \phi$. Such a coupling would arise from the conformal transformation \eqref{conformal_transformation} if we add matter in the Jordan frame. The value of $C$ is then constrained by observations, but these constraints can be alleviated if the quintessence field only couples to dark matter \cite{Damour:1990tw}, or if $C$ is time-dependent \cite{Amendola:1999er}. For simplicity, we did not include this coupling in the present analysis, leaving it for future work.}

\vfill

Let us now assume a FLRW background with scale factor $a(t)$ for the metric and write the equations {of motion} in terms of the Hubble rate $H(t) = \dot a/a$, the matter energy density $\rho(t)$ and the the scalar field $\phi(t)$. The Lagrangian in (\ref{action_scalar_field}) is of the form
\be {\cal{L}}_\phi = -\frac{1}{2}F(\phi)(\nabla\phi)^2-U(\phi)+\frac{1}{4}G(\phi)(\nabla\phi)^4,\ee 

leading to the following scalar e.o.m.
\be\label{EOMS}
(F+3G\dot{\phi}^2)\ddot{\phi}+3H\dot{\phi}(F+G\dot{\phi}^2) + \frac{1}{2}F'\dot{\phi}^2
+\frac{3}{4}G'\dot{\phi}^4+U'=0,
\ee

\vfill\newpage\vfill

along with the Friedmann equations
\be\label{friedmann_equations}
H^2 = \frac{1}{3}(\rho_{\phi} + \rho) \hspace{\baselineskip}\text{and}\hspace{\baselineskip}
\dot{H} = -\frac{1}{2}(\rho_{\phi}+p_{ \phi}) - \frac{1}{2}(1+w)\rho,
\ee

where
\be
\rho_{ \phi} = \frac{1}{2}F\dot{\phi}^2+U+\frac{3}{4}G\dot{\phi}^4
\hspace{\baselineskip}\text{and}\hspace{\baselineskip}
p_{ \phi} = \frac{1}{2}F\dot{\phi}^2-U+\frac{1}{4}G\dot{\phi}^4.
\ee

Inserting the explicit form of the functions $F, G, U$ in the previous equations, we end up with
\begin{eqnarray}
\ddot\phi(1+\xi\phi^2 + 3\alpha\dot\phi^2) + 3H\dot\phi(1+\xi\phi^2+\alpha\dot\phi^2) + V' + \xi\phi\dot\phi\nonumber\\
-\frac{\xi\phi(1+\xi\phi^2)+\alpha V'}{(1+\xi\phi^2)^2+4\alpha V}\left(2(1+\xi\phi^2)\dot\phi^2 + 4V+3\alpha\dot\phi^4\right) = 0,\label{equation_of_motion}
\end{eqnarray}

and
\begin{equation}\label{DENSITIES}
\rho_\phi = \frac{1/2(1+\xi\phi^2)\dot\phi^2 + V + 3/4\alpha\dot\phi^4}{(1+\xi\phi^2)^2 + 4\alpha V}, 
\hspace{\baselineskip}
p_\phi = \frac{1/2(1+\xi\phi^2)\dot\phi^2 - V + 1/4\alpha\dot\phi^4}{(1+\xi\phi^2)^2 + 4\alpha V}.
\end{equation}

\vspace{\baselineskip}

{The system would be much different in the metric formulation, because in that case the quadratic correction $\alpha R^2$ introduces a second scalar degree of freedom, resulting in a two-fields dynamics that is not in the scope of our analysis. In the absence of the $\alpha R^2$ term, the non-minimal coupling $\xi\phi^2 R$ has been studied in the metric formalism, e.g. in \cite{Uzan:1999ch, Chiba:1999wt}.}

\vspace{\baselineskip}

In the rest of our paper, we study the system \eqref{friedmann_equations}-\eqref{equation_of_motion} for an exponential potential $V(\phi) = M^4\exp(-\lambda\phi)$, a simple choice well-motivated by string theory and supergravity models. More precisely, we investigate whether this system possesses ``realistic" solutions that realize the sequence {\textit{radiation domination}} $\rightarrow$ {\textit{matter domination}} $\rightarrow$ {\textit{accelerated expansion}} with the correct duration and the scalar field stays subdominant during the two first eras and the standard evolution of the Universe (e.g. nucleosynthesis) is not disturbed. For this, we use dynamical system methods, that we review in passing. We start by considering the quadratic correction $\alpha R^2$, then the nonminimal coupling $\xi\phi^2 R$, and proceed {to analyze} the combination of both.

\vfill\newpage\vfill

\section{Quadratic correction}

\vfill

Let us start by considering the case $\xi = 0$, where the only non-standard term in (\ref{starting_point_action}) is the quadratic correction, or Starobinsky term $\alpha R^2$. In this case, we see in (\ref{action_scalar_field}) that the dynamics is essentially the one of a noncanonical scalar field in a potential $U = V/(1+4\alpha V)$. The new -phenomenological- potential exhibits a plateau ($U \simeq 1/(4\alpha)$ whenever $4\alpha V \gg 1$) on which primordial slow-roll inflation can take place. In the case of quintessence with an exponential potential $V(\phi) = M^4\exp(-\lambda\phi)$, the phenomenological potential $U$ has this plateau for $\phi\rightarrow-\infty$, and an exponential tail for $\phi\rightarrow+\infty$. It is known that accelerated expansion appears as the future attractor in an exponential potential with $\lambda^2 < 2$ \cite{BBCC}. In this case, the past attractor corresponds to a Universe dominated by the scalar field kinetic energy with $w=1$ and the matter dominated Universe only corresponds to a saddle point. This undesirable past attractor is most often ignored, as one can always argue that the system starts away from it and already close to the radiation/matter dominated point after inflation. We will show that the quadratic correction $\alpha R^2$ can modify this behavior. {The new $4\alpha V \gg 1$ region of the dynamical system contains two consecutive saddle points that respectively correspond to matter domination and accelerated expansion; it can thus reproduce the evolution of the Universe. However,} compatibility with the present day {Hubble rate} depends on the value of the parameter $\alpha$, {and yields} an unacceptably large value for $\alpha$. As a result, for reasonable values of $\alpha$, the $\alpha R^2$ term does not play much of a role in the late time evolution. Still, despite this negative result, employing dynamical systems methods to study this case {will allow us to review them before studying the case with $\xi \neq 0$} that leads to a different result.

\vfill

When $\xi = 0$, equations (\ref{equation_of_motion}) and (\ref{friedmann_equations}), along with (\ref{DENSITIES}) become the following
\be\label{EOMS-1}
\ddot{\phi}(1+3\alpha\dot{\phi}^2)+3H\dot{\phi}(1+\alpha\dot{\phi}^2)
+\frac{V'}{1+4\alpha V}(1-2\alpha\dot{\phi}^2-3\alpha^2\dot{\phi}^4) = 0,
\ee  

and
\begin{equation}\label{friedmann_equations_alpha}
H^2 = \frac{2\dot\phi^2 + 3\alpha\dot\phi^4 + 4V}{12(1+4\alpha V)} + \frac{\rho}{3}
\hspace{\baselineskip}\text{and}\hspace{\baselineskip}
\dot H = -\frac{\dot\phi^2(1+\alpha\dot\phi^2)}{2(1+4\alpha V)} - \frac{1}{2}(1+w)\rho.
\end{equation}

\vfill

The starting point of the analysis is to carefully choose dynamical variables in order to rewrite (\ref{EOMS-1}) and (\ref{friedmann_equations_alpha}) as an autonomous system of first order equations. This can be achieved with
\be x \equiv \frac{\dot{\phi}}{2\sqrt{V}},\hspace{\baselineskip}
y\equiv\frac{\dot{\phi}}{2H},\hspace{\baselineskip}
\lambda \equiv-\frac{V'}{V}\hspace{\baselineskip}\text{and}\hspace{\baselineskip}
\nu\equiv4\alpha V.
{\label{VARIABLES}}\ee

Here, we assumed that $V > 0$. Furthermore, as we said earlier, we restrict our analysis to an exponential potential, for which $\lambda$ is constant. These are not the only variables that one can choose to analyze the system (\ref{EOMS-1}), (\ref{friedmann_equations_alpha}), but these make it simple in the $\nu\gg 1$ regime. 

\vfill\newpage\vfill

These new variables obey the following autonomous system, where the role of time is played by the {\textit{number of e-folds}} $N\sim\ln a$, or $dN=Hdt$
\begin{equation}\label{equation_alpha_x}
\frac{dx}{dN} = \frac{\lambda y-6(1+\nu)x^2(1+\nu x^2)+\lambda x^2y(2+3\nu(2+\nu)x^2)}{2(1+\nu)x(1+3\nu x^2)},
\end{equation}

and
\begin{eqnarray}
\frac{dy}{dN} &=& -\frac{y}{2(1+\nu)x^2(1+3\nu x^2)}\nonumber\\
&\times&\left\{y((1+w)y-\lambda) - 3(1-3w)\nu^2x^6y^2\right.\nonumber\\
&-&\nu x^4(3+3\nu-3\lambda\nu y+7y^2 + 9w(1+\nu-y^2))\nonumber\\
&+&\left. x^2((1-w)(3-2y^2)+\nu(2\lambda y + 3(1+y^2)+3w(y^2-1)))\right\},{\label{equation_alpha_y}}
\end{eqnarray}

and finally
\begin{equation}\label{equation_alpha_nu}
\frac{d\nu}{dN} = -2\lambda\nu y.
\end{equation}

For non exponential potentials, there would also be an equation for $\lambda$. The system is invariant under $x\rightarrow\,-x$, so we can restrict ourselves to $x>0$ and obtain trajectories with $x<0$ by reflection. The system is also invariant under $(\lambda, y) \rightarrow (-\lambda, -y)$, so we can restrict ourselves to $\lambda>0$. Furthermore, $x>0$ implies $\dot{\phi}>0$. Then, we would have $y<0$, if and only if $H<0$, as seen in (\ref{VARIABLES}). {So} restricting ourselves to expanding cosmologies implies that $y>0$. Since $dy/dN=0$ when $y=0$, the system cannot dynamically cross $y=0$ and $y$ remains positive during the entire evolution. Assuming $\alpha>0$ also implies $\nu>0$. So in the end, all variables are positive. Two quantities of interest are the scalar field energy density parameter $\Omega_\phi = \rho_\phi/H^2$ and the equation of state parameter $w_\phi = p_\phi/\rho_\phi$. They can be expressed as
\begin{equation}\label{cosmological_parameters_alpha}
\Omega_\phi = \frac{y^2(1+2x^2+3\nu x^4)}{3x^2(1+\nu)}
\hspace{\baselineskip}\text{and}\hspace{\baselineskip}
w_\phi = -\frac{1-2x^2-\nu x^4}{1+2x^2+3\nu x^4}.
\end{equation}

The values taken by the $(x, y, \nu)$ are constrained by $\Omega_\phi < 1$, otherwise one would need $\rho < 0$. Let us also define the effective equation of state parameter $w_{\mathrm{eff}} = w_\phi\Omega_\phi + w(1-\Omega_\phi)$.

\vfill

The system of equations (\ref{equation_alpha_x})-(\ref{equation_alpha_nu}) looks complicated, but we are only interested in its fixed points, that is, the points $\bm{P} = (x_*, y_*, \nu_*)$ of the phase space where $d\bm{P}/dN=0$, because they encapsulate the qualitative evolution of the system. And in particular, the equation $d\nu/dN = 0$ is simply solved by $\nu_* = 0$ or $y_* = 0$. We can also have $\nu \rightarrow +\infty$ in such a way that $d\nu/dN \rightarrow 0$. Indeed if we let $\tilde\nu \equiv \nu/(1+\nu)$, we get
\be\label{equation_nu_tilde}
\frac{d\tilde{\nu}}{dN}=2\lambda\,\tilde{\nu}(\tilde{\nu}-1)y,
\ee

so $d\tilde\nu/dN = 0$ when $\tilde{\nu}=0$ or $1$, corresponding to $\nu = 0$ and $\nu \rightarrow+\infty$. Since $(y, \nu) > 0$, (\ref{equation_alpha_nu}) implies that $\nu$ decreases with $N$. So $\nu\rightarrow+\infty$ (resp. $0$) corresponds to the past (resp. future). 

\vfill\newpage\vfill

Starting with {\textit{early times}} {with} $\nu\rightarrow+\infty$, the system (\ref{equation_alpha_x})-(\ref{equation_alpha_y}) becomes
\be
\frac{dx}{dN}=\frac{x}{2}(\lambda y-2)
\hspace{\baselineskip}\text{and}\hspace{\baselineskip}
\frac{dy}{dN}=\frac{y}{2}\left[(1-3w)x^2y^2-\lambda y+(1+3w)\right],
{\label{EQS-3}}\ee

while the cosmological parameters (\ref{cosmological_parameters_alpha}) are
\be \label{cosmological_parameters_alpha_early}
\Omega_{ \phi}\simeq x^2y^2
\hspace{\baselineskip}\text{and}\hspace{\baselineskip}w_{ \phi}\simeq\frac{1}{3}.\ee

Let us first notice that in the case of a radiation background fluid ($w=1/3$), the system (\ref{EQS-3}) simplifies further and can be solved easily to give
\be x(N)=\frac{x_0(2+\lambda y_0(e^N - 1))}{2e^N}
\hspace{\baselineskip}\text{and}\hspace{\baselineskip}
y(N)=\frac{2y_0 e^N}{2 + \lambda y_0(e^N - 1)},{\label{X0Y0}}.\ee

In particular, the scalar field energy density (\ref{cosmological_parameters_alpha_early}) is constant; namely, $\Omega_\phi \simeq x^2y^2 = x_0^2y_0^2$. Therefore, if the scalar field is subdominant initially, to satisfy for instance the nucleosynthesis constraint, it shall remain subdominant during all of the radiation dominated era. The transition to matter domination then proceeds as usually without being affected by the presence of the scalar field. 

\vfill

In the case of a matter background fluid ($w=0$), the analysis is not much more complicated. We give the fixed points of the system along with their properties in the following table:
\begin{table}[ht]
\begin{center}\begin{tabular}{c||c|c|c||c|c|c}
Point & $x_*$ & $y_*$ & eigenvalues & $w_{\mathrm{eff}}$ & $\Omega_\phi$ & stability \\
\hline\hline
$O$ & $0$ & $0$ & $-1, 1/2$ & 1/3 & $0$ & saddle\\
$A$ & $\lambda/2$ & $2/\lambda$ & $-1, 1$ & 1/3 & $1$ & saddle\\
$B$ & $0$ & $1/\lambda$ & $-1/2, -1/2$ & 1/3 & $0$ & attractive
\end{tabular}
\caption{fixed points of the system (\ref{EQS-3}) that are not at infinity.}
\label{fixed_points_alpha_early}
\end{center}
\end{table}

Here, the eigenvalues are the ones of the Jacobian matrix of the system around the corresponding fixed points. According to linear stability theory, the point is stable, or attractive (resp. unstable, or repulsive) if the real part of its eigenvalues are negative (resp. positive), and saddle if some are negative and some positive. If some of the eigenvalues have vanishing real parts, we need to go beyond linear stability analysis to assess the stability of the point. 

\vfill

In our case, $B$ is the only stable point and is thus the future attractor of the system. This is good, because $B$ corresponds to a matter dominated Universe, which is what we should get at early times. In that respects, we could worry that the system gets too close to the scalar dominated saddle point $A$ but we can show from (\ref{cosmological_parameters_alpha_early}) and (\ref{EQS-3}) that $d\Omega_\phi/dN = \Omega_\phi(\Omega_\phi - 1) < 0$, so $\Omega_\phi$ only decreases. Once again, if the scalar field is subdominant initially, it remains subdominant during all of the matter dominated era or at least as long as $\nu \gg 1$.

\vfill\newpage\vfill

The constraint $\Omega_\phi < 1$ does not make the phase space compact; it is possible to have fixed points where one of $x$ or $y$ goes to infinity. To study them, we make a change of variable analogous to the one we made in (\ref{equation_nu_tilde}); namely, $\tilde x = x/(1+x)$ and $\tilde y = y/(1+y)$, and obtain a new system. This new system has divergences, e.g.
\begin{equation}\label{equation_xtilde_alpha}
\frac{d\tilde x}{dN} = -\frac{\tilde x(\tilde x-1)(2-(2+\lambda)\tilde y)}{2(\tilde y-1)},
\end{equation}

and similarly for $d\tilde y/dN$. These divergences can be handled knowing that for any dynamical system $d\bm P/dN = \bm f(\bm P)$, the system $d\bm P/dN = \bm \xi(\bm P)\bm f(\bm P)$ for positive definite $\xi(\bm P) > 0$ has the same fixed points with the same properties. Using this trick in our case, we end up with two new fixed points:

\begin{center}\begin{tabular}{c||c|c|c||c|c|c}
Point & $\tilde x_*$ & $\tilde y_*$ & eigenvalues & $w_{\mathrm{eff}}$ & $\Omega_\phi$ & stability \\
\hline\hline
$P_x$ & $1$ & $0$ & $0, 0$ & 1/3 & $-$ & repulsive\\
$P_y$ & $0$ & $1$ & $\lambda, \lambda$ & 1/3 & $-$ & repulsive\\
\end{tabular}\end{center}

We cannot conclude about the stability of $P_x$ using its eigenvalues, but we can use numerical computations to asses that it is repulsive. For both $P_x$ and $P_y$, the energy density $\Omega_\phi = \tilde x^2\tilde y^2/((1-\tilde x)^2(1-\tilde y)^2)$ is undetermined but can be found to be $\Omega_\phi = 1$ numerically. In any case, as we said earlier, in a realistic evolution, $\Omega_\phi$ would have to be constant and small in the radiation dominated era and it would only decrease from there during matter domination.

\vfill

Let us now analyze the main consequence of $\nu$ actually being finite. Since the future attractor $B$ has $x_* = 0$, even if $\nu$ is large we will have $\nu x^4 \ll 1$ at some point. When this happens, equation (\ref{cosmological_parameters_alpha_early}) becomes
\begin{equation}\label{cosmological_parameters_alpha_early_simplified}
\Omega_\phi \simeq \frac{y^2}{3\nu x^2}
\hspace{\baselineskip}\text{and}\hspace{\baselineskip}w_{ \phi}\simeq-1.
\end{equation}

\vfill

At the double limit $\nu \gg 1$ and $x \ll 1$ with $\nu x^2 \sim 1$, the system of (\ref{equation_alpha_x})-(\ref{equation_alpha_y}) becomes
\begin{eqnarray}\label{system_alpha_early_simplified}
\frac{dx}{dN}&=&\frac{x}{2}(\lambda y-2)+\frac{1}{\nu x}\left(\frac{\lambda y}{6\nu x^2}-1\right),\\
\frac{dy}{dN}&=&\frac{y}{2}\left(1-\lambda y+\frac{y(\lambda-y)}{\nu x^2}-\frac{4(1+\lambda y)}{(1+3\nu x^2)}\right).
\end{eqnarray}

Here we consider $w = 0$ for simplicity. This system has one real fixed point, but its expression is too complicated to be insightful. A further simplification we can make is to neglect the first term $x/2(\lambda y-2)$ in (\ref{system_alpha_early_simplified}) using $x \ll 1$. This is not strictly correct, because the system would then be attracted to a fixed point that cancels the second term only. But if the second term vanishes, the first one is not negligible anymore. If do it anyway, we obtain
\begin{equation}\label{WOULDBE}
x_*^T \simeq \sqrt{\frac{\lambda y_*^T}{6\nu}}, \hspace{\baselineskip}\text{and}\hspace{\baselineskip}
y_*^T \simeq \frac{12\lambda}{12+3\lambda^2+\sqrt{3(48+72\lambda^2+11\lambda^4)}}.
\end{equation}

\vfill

The point (\ref{WOULDBE}) can be shown to be stable. It is the one to which the system is attracted to instead of $B$ when $\nu x^2 \sim 1$. It depends on $\nu$, so it is not really a fixed point, but what we can call a \textit{tracked point}. Since the system lags behind this point as it moves, the expressions (\ref{WOULDBE}) cannot be used to describe its behavior very precisely. In particular, using (\ref{WOULDBE}) with (\ref{cosmological_parameters_alpha_early_simplified}) gives $\Omega_\phi \simeq 2y_*^T/\lambda$ which is not the $\Omega_\phi \simeq 1$ that we will later observe numerically.

\vfill

\begin{figure}[ht]\centering\includegraphics[scale=.8]{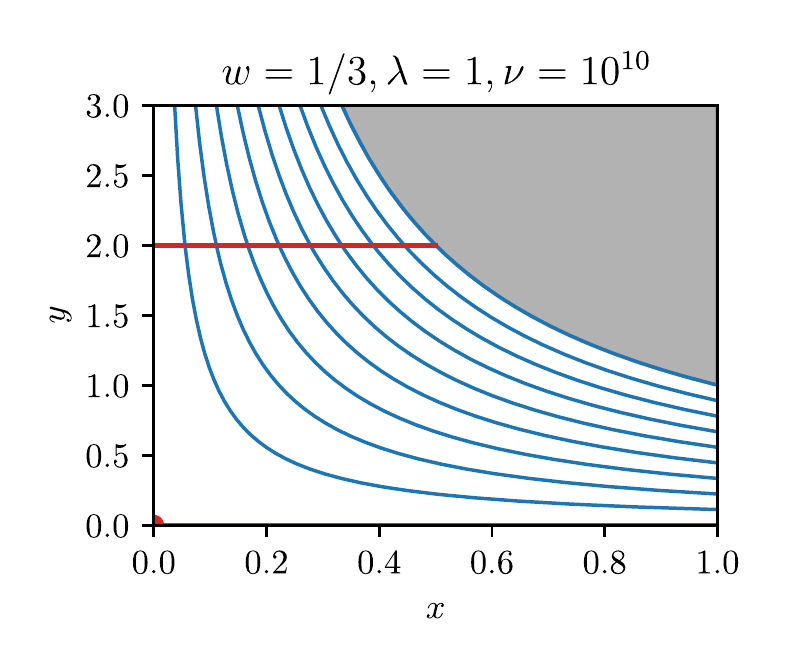}\\
\includegraphics[scale=.8]{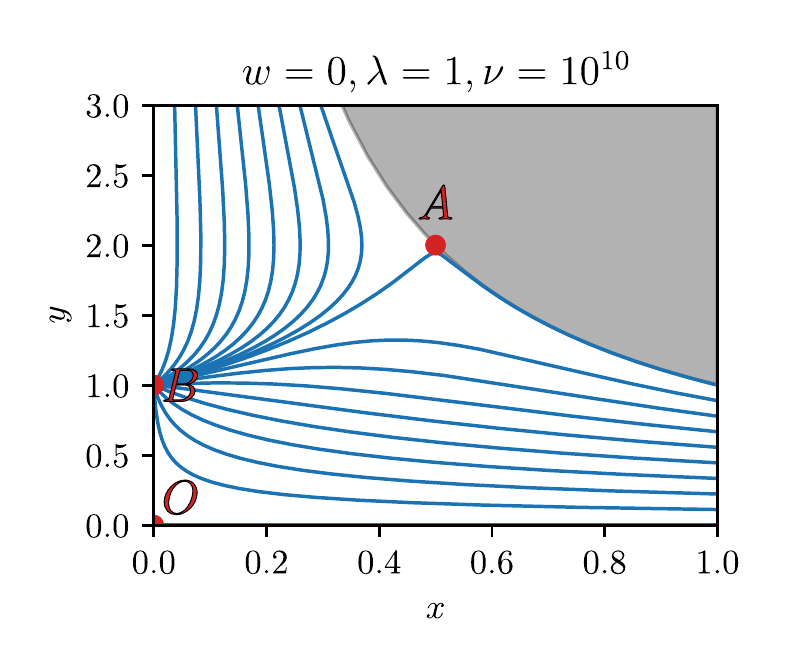}
\includegraphics[scale=.8]{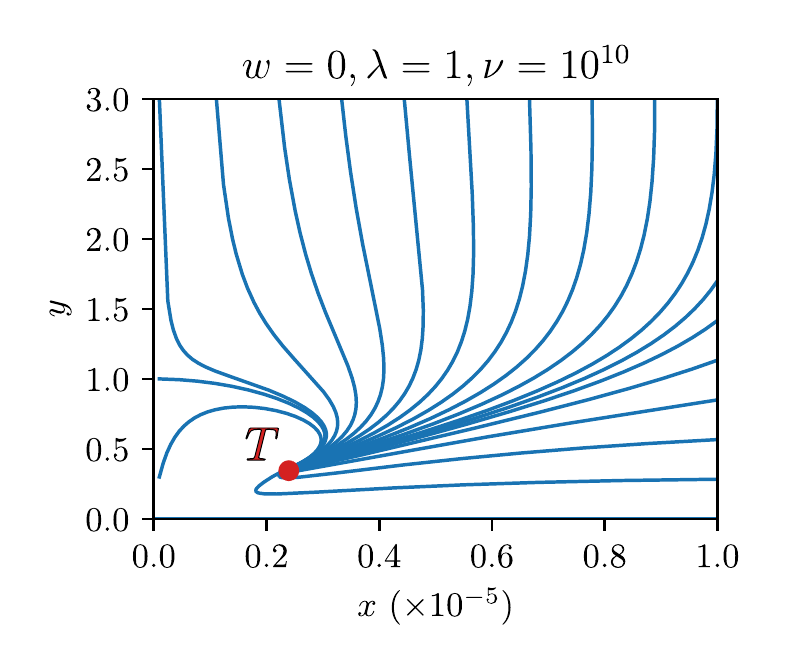}
\caption{The phase space of (\ref{equation_alpha_x})-(\ref{equation_alpha_y}) with $\nu = 10^{10}, \lambda = 1$, $w = 1/3$ (top) $w=0$ (bottom). The right panel in the bottom is a zoom on the left one for small $x$. Fixed points are indicated and labeled as in Table \ref{fixed_points_alpha_early}. The shaded region is the one excluded by the positive energy constraint $\Omega_\phi < 1$.}
\label{phase_space_alpha}
\end{figure}

In Figure \ref{phase_space_alpha}, we plot the phase space of the system (\ref{equation_alpha_x})-(\ref{equation_alpha_y}) for a constant $\nu \gg 1$ and generic value of $\lambda$. In the case $w = 1/3$, the trajectories are indeed lines of constant $\Omega_\phi = x^2y^2$ as found in (\ref{X0Y0}). They all converge to the fixed line $y = 2/\lambda$. The origin is a saddle point. In the case $w = 0$, we check the presence of the fixed points listed in Table \ref{fixed_points_alpha_early}, manifestly with the right properties. The behavior at infinity also corresponds to the one discussed earlier. Since $\nu$ is finite in this plot, $B$ is not the true future attractor. When $x$ gets small the system is in fact attracted by the point $T$ approximately located at coordinates (\ref{WOULDBE}). As $\nu$ decreases, this point moves towards higher values of $x$ until the approximation $\nu \gg 1$ starts to break.

\vfill

\vfill\newpage\vfill

At late times, $\nu\rightarrow0$, the scalar field is in the tail of the phenomenological potential $U = V/(1+4\alpha V)$ and we recover the well known quintessence in an exponential potential. In this limit, we could perform the same analysis as we just did with the variables ({\ref{VARIABLES}}). Since $\nu\rightarrow0$ corresponds to late times in our framework, let us just remind what is its future attractor. When $\lambda^2 < 3$, it corresponds to a scalar dominated Universe with $w_\phi = \lambda^2/3 - 1$. In this case, the experimental $-1 < w_{ \phi} < -0.95$ implies that $\lambda<0.34$ {(in \cite{Montefalcone:2020vlu}, authors obtain a larger bound $\lambda < 0.6$ by taking into account the evolution of $w_\phi$)}. If $\lambda^2>3$, the late time attractor is a scaling solution with $\Omega_{ \phi}=3/\lambda^2$ and $w_{ \phi}=w$ and there is no accelerated expansion.

\vfill

\begin{figure}[ht]\begin{center}
\includegraphics[scale=.8]{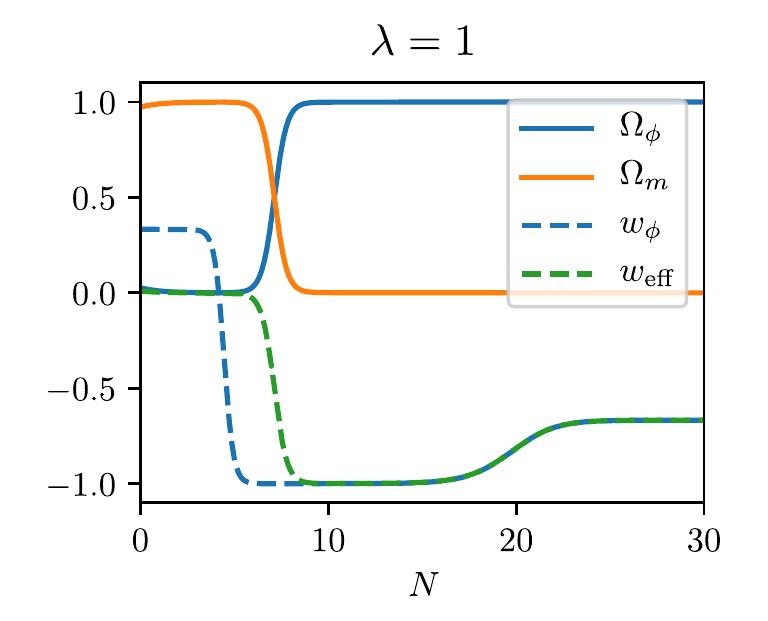}
\includegraphics[scale=.8]{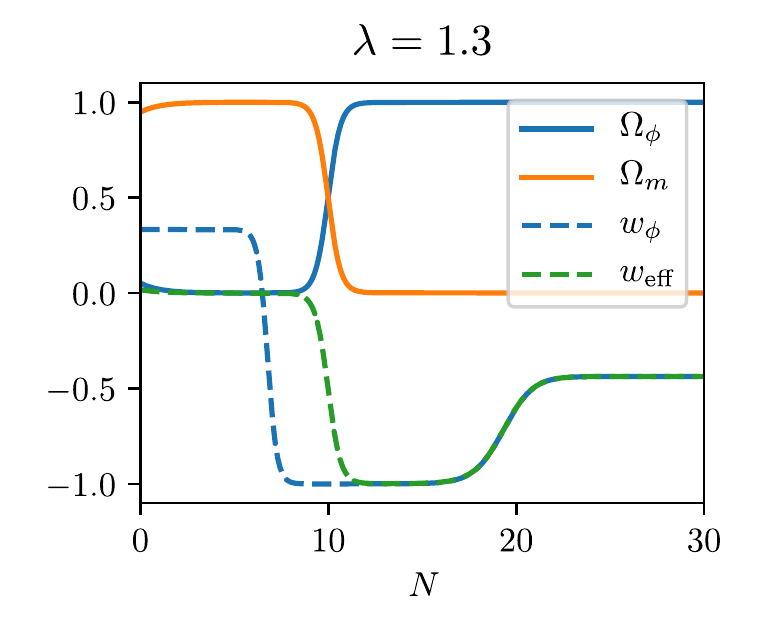}\\
\includegraphics[scale=.8]{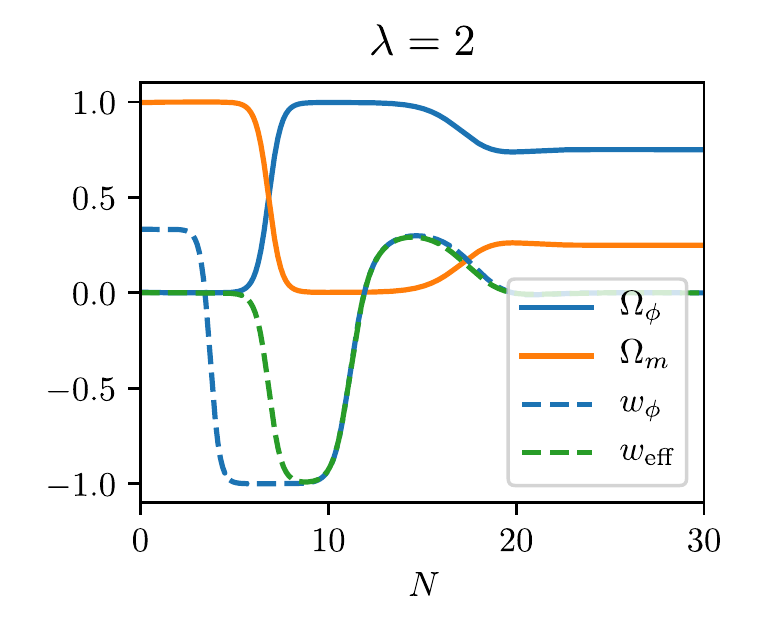}
\includegraphics[scale=.8]{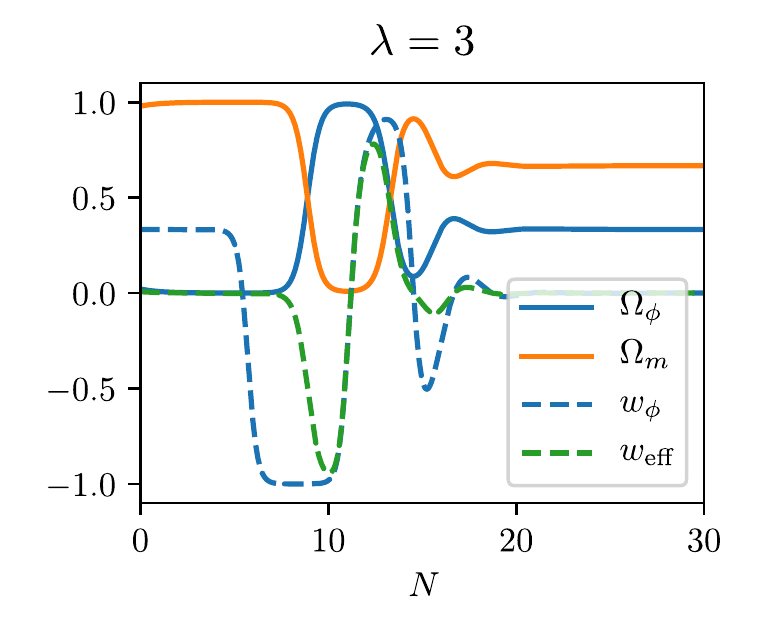}\end{center}
\caption{Evolution of the scalar field parameters (\ref{cosmological_parameters_alpha}) following from the system (\ref{equation_alpha_x})-(\ref{equation_alpha_nu}), for $w = 0$ and different values of $\lambda$. We denoted $\Omega_m = 1-\Omega_\phi$ and $w_{\mathrm{eff}} = w\Omega_m + x_\phi\Omega_\phi$. We choose initial $\nu_i = 10^{10}$ to have between $8$ and $10$ e-folds of matter domination and $x_i, y_i$ to have $\Omega_{ \phi}<0.2$ initially.}
\label{evolution_alpha}
\end{figure}

The evolution of the scalar field parameters (\ref{cosmological_parameters_alpha}) following from the system (\ref{equation_alpha_x})-(\ref{equation_alpha_nu}), is shown in Figure \ref{evolution_alpha} for different values of $\lambda$. In all cases, we observe similar evolutions that confirm the previous analysis. Initially, $\nu\gg 1$ so (\ref{cosmological_parameters_alpha_early}) applies and $w_\phi \simeq 1/3$. The system is attracted towards point $B = (0, 1/\lambda)$ of Table \ref{fixed_points_alpha_early} and quickly $\Omega_\phi \simeq 0$. Then, as $x\rightarrow0$, the other approximation (\ref{cosmological_parameters_alpha_early_simplified}) applies and $w_\phi \simeq -1$. The system tracks the moving point $T$ and it turns out that as it does so $\Omega_\phi \simeq 1$. Finally, as $\nu$ keeps on decreasing following (\ref{equation_alpha_nu}), we end up with $\nu\simeq 0$ where we recover the usual quintessence behavior that we just reviewed.

\vfill

\vfill\newpage\vfill

An interesting feature of this evolution is the presence of an accelerated expansion era, with $w_\phi \simeq -1$, between matter-scalar equality and the usual quintessence era. As mentioned earlier, this corresponds to when the system is tracking the point $T$ of Figure \ref{phase_space_alpha}. This era only gets shorter as $\lambda$ increases, effectively vanishing beyond $\lambda \simeq 3$. Its presence alleviates the bound $\lambda < 0.34$ that ensured $w_\phi \simeq -1$ for the usual quintessence in an exponential potential.

\vfill

Up to now, we have shown that, if $\lambda < 3$, the system (\ref{equation_alpha_x})-(\ref{equation_alpha_nu}) can lead to a realistic background evolution of the Universe, with $\nu_i$ tunable to obtain between $8$ and $10$ e-folds of matter domination, followed by accelerated expansion with $w_{\mathrm{eff}}\sim -1$. We now discuss the value of $\alpha$. Numerically, we can extract the values $(x_{\mathrm{eq}}, y_{\mathrm{eq}}, \nu_\mathrm{eq})$ at matter-scalar equality. Since this approximately happened today, the Hubble rate at this point is $H_{\mathrm{eq}}=H_0=1.46\times 10^{-42}\ \mathrm{GeV}$. Then, from (\ref{VARIABLES}), reintroducing $\kappa=1.1\times 10^{-19}\ \mathrm{GeV}^{-1}$, we have
\be \alpha = \frac{\nu_{\mathrm{eq}}x_{\mathrm{eq}}^2}{4 y_{\mathrm{eq}}^2}\frac{1}{(\kappa H_0)^2}\simeq \frac{\nu_{\mathrm{eq}}x_{\mathrm{eq}}^2}{4 y_{\mathrm{eq}}^2}\times10^{121}\,.\ee

As we see, the smallness of $H_0$ implies the huge factor $10^{120}$ in this equation that implies a huge value for the parameter $\alpha$ unless $\nu_{\mathrm{eq}}x_{\mathrm{eq}}^2/(4 y_{\mathrm{eq}}^2) \ll 1$. But this cannot happen for a simple reason. As we have seen, matter-scalar equality happens when the system starts seeing and tracking the point $T$. Even if the formula for $\Omega_\phi$ in (\ref{cosmological_parameters_alpha_early_simplified}) is approximate, we can still use it to evaluate that 
\begin{equation}
\Omega_{\phi, \mathrm{eq}} = \frac{y_{\mathrm{eq}}^2}{3\nu_{\mathrm{eq}} x_{\mathrm{eq}}^2} \sim 0.5-0.7 \hspace{\baselineskip}\rightarrow\hspace{\baselineskip}
\frac{\nu_{\mathrm{eq}}x_{\mathrm{eq}}^2}{4 y_{\mathrm{eq}}^2}\sim 0.1.
\end{equation}

which is not much smaller than $1$. So there is no way to alleviate a huge value $\alpha \sim 10^{120}$, {with the Hubble rate having its value today}. We have verified this numerically. This ridiculous value comes from trying to make the $\alpha R^2$ term play a major role in the dynamics. Reversing this conclusion, it is safe to say that for reasonable values of $\alpha$, the $\alpha R^2$ term never plays any role in the late time evolution, at least in the Palatini formulation. This is in contrast with the non-minimal coupling $\xi\phi^2R$, the effects of which will be analyzed in the next sections. 

\vfill

{Closing this one, we may give a few words on the link between our work and quintessential inflation in the Palatini formulation \cite{D1, D2, V}. In quintessential inflation, the scalar field is responsible for inflation both at primordial and current times. In \cite{D1, D2, V}, primordial inflation happens with negligible $\alpha(\partial\phi)^4$ and $4\alpha V \gg 1$. This corresponds to when the system is tracking the point $T$ in our analysis, and in this period we indeed obtain accelerated expansion. The key difference is that we identified it with the late time instead of early time inflation; in quintessential inflation, late time inflation can be obtained via other features of the potential. Note also that in this case matter is not present before reheating, so anything that precedes the point $T$ in our analysis does not apply. Finally, note that contrary to what happens in generic quintessential inflation, we do not have an era of kination ($\Omega_\phi = 1, w_\phi = 1$) after inflation (see Figure \ref{evolution_alpha}). This may be due to the absence of reheating in our case, since current inflation happens after the era of matter domination.}

\vfill\newpage\vfill

\section{Non-minimal coupling}\label{sec:non-minimal_coupling}

\vfill

We now consider the case $\alpha = 0$, where the only non-standard term in (\ref{starting_point_action}) is the non-minimal coupling $\xi\phi^2R$. In this case, we could start as before from equations (\ref{equation_of_motion}) and (\ref{friedmann_equations}), analyzing them with carefully chosen dynamical variables. A good choice seems to be
\be X^2\equiv\frac{\dot{\phi}^2}{6H^2(1+\xi\phi^2)}, \hspace{\baselineskip} 
Y^2\equiv\frac{V}{3H^2(1+\xi\phi^2)^2}, \hspace{\baselineskip} 
Z^2\equiv\phi^2 \hspace{\baselineskip}\text{and}\hspace{\baselineskip}
\lambda\equiv-\frac{V'}{V},{\label{VARIABLES-1}}\ee

because in this case $\Omega_\phi = X^2+Y^2$. However, we can proceed in a more intuitive way if we rewrite the action in terms of a canonical scalar. Indeed, starting from (\ref{action_scalar_field}) and introducing
\be  \Phi=\int\frac{d\phi}{\sqrt{1+\xi\phi^2}}=\frac{1}{\sqrt{\xi}}\sinh^{-1}\left(\sqrt{\xi}\phi\right)
\hspace{\baselineskip}\Longleftrightarrow\hspace{\baselineskip}
\phi=\frac{1}{\sqrt{\xi}}\sinh\left(\sqrt{\xi}\Phi\right),{\label{FI}}\ee 

the action for $\Phi$ is

\be\label{action_alpha_nonminimal}
{\cal{S}}_\Phi = \int d^4x\sqrt{-g}\left\{
-\frac{1}{1+4\alpha U(\Phi)}\left(\frac{1}{2}(\nabla\Phi)^2 + U(\Phi) - \frac{\alpha}{4}(\nabla\Phi)^4\right)\right\},
\ee

where in the case of an exponential potential
\be U(\Phi)=\frac{V(\phi(\Phi))}{(1+\xi\phi(\Phi)^2)^2}=\frac{M^4\exp({-\lambda/\sqrt{\xi}\sinh(\sqrt{\xi}\Phi)})}{\cosh^4(\sqrt{\xi} \Phi)}.{\label{EFPOT}}\ee

In the case $\alpha = 0$, the action (\ref{action_alpha_nonminimal}) boils down to a canonical scalar field in the \textit{phenomenological potential} $U$, which contains all the $\xi$ dependence. We can thus understand most of the dynamics just by studying the behavior of $U$. When $\Phi \rightarrow -\infty$, it goes very quickly to $+\infty$, while when $\Phi\rightarrow +\infty$, it goes very quickly to $0$. By ``very quickly", we mean faster than exponential. This behavior is due to the $\exp(-\lambda/\sqrt{\xi}\sinh(\sqrt{\xi}\Phi))$ in (\ref{EFPOT}), which is like a {\textit{``double exponential"}} in these limits. As we remarked earlier, a regular exponential $\exp(-\lambda\phi)$ with $\lambda^2 > 3$ is already too steep to provide accelerated expansion and, therefore, we do not expect it to happen in the tail $\Phi\rightarrow +\infty$ of $U$ either. We shall show that it does not. In between these limits, if $4\xi > \lambda^2$, the phenomenological potential has two local extrema, located at (note that this is in terms of the non canonical scalar)
\be \phi_{\mathrm{min/max}}=-\frac{2}{\lambda}\left(1\pm\sqrt{1-\frac{\lambda^2}{4\xi}}\right).\ee 

The corresponding values of $U$ are
\be
U_{\mathrm{min/max}}=\frac{\lambda^4M^4}{64\xi^2}\frac{\exp(2\pm2\sqrt{1-\lambda^2/(4\xi)})}{(1\pm\sqrt{1-\lambda^2/(4\xi)})^2}\,.
{\label{EXTREMA}}\ee 

When $4\xi \gg \lambda^2$, this becomes
\begin{equation}\label{potential_U_limits}
U_{\mathrm{min}}\simeq\frac{\exp(4)\lambda^4M^4}{256\xi^2}
\hspace{\baselineskip}\mathrm{and}\hspace{\baselineskip}
U_{\mathrm{max}}\simeq M^4
\end{equation}

\vfill\newpage\vfill

We will show that the system has two future attractors: one where the field ends up in the local minimum, and another where it rolls down the tail. The choice between the two essentially depends on whether the field starts on one side or the other of the local maximum. The first attractor is the only one that corresponds to accelerated expansion. There the field acts as a cosmological constant $\Lambda_{\mathrm{eff}}\sim U_{\mathrm{min}}$. As we see in (\ref{potential_U_limits}), the parameters $\lambda, \xi$ play a role in $\Lambda_{\mathrm{eff}}$ and can help lowering it. To reproduce the current value of $H_0$, we need
\begin{equation}\label{lambda_M_xi_nonminimal}
\frac{\lambda^2M^2}{\xi}\sim 10^{-23}\ \mathrm{GeV}^2\,.
\end{equation}

Note that in the region where $\sqrt{\xi}\Phi\ll 1$, the phenomenological potential $U$ in (\ref{EFPOT}) is approximated by the usual exponential $U(\Phi) \simeq M^4\exp(-\lambda\Phi)$. Therefore, the future attractors of this usual exponential discussed earlier can be approached in a transient way.

\vfill

Let us start the analysis. It is very well known that the equations stemming from (\ref{action_alpha_nonminimal}) (with $\alpha=0$) can be studied using the so-called \textit{expansion normalised variables} \cite{BBCC}
\be {\label{VARIABLES-2}}
x\equiv\frac{\dot{\phi}}{\sqrt{6}H},
\hspace{\baselineskip} 
y\equiv\frac{1}{H}\sqrt{\frac{V}{3}},
\hspace{\baselineskip}\text{and}\hspace{\baselineskip}\Lambda\equiv-\frac{U'}{U},\ee 

in terms of which
\be \label{scalar_parameters_nonminimal}
\Omega_{ \phi} = x^2+y^2,
\hspace{\baselineskip}\text{and}\hspace{\baselineskip}
w_{ \phi}=\frac{x^2-y^2}{x^2+y^2}.\ee

{and} the system obeys the following system of equations
\begin{eqnarray}
\frac{dx}{dN}&=&-\frac{3}{2}\left(2x+(w-1)x^3+(w+1)x(y^2-1)-\sqrt{\frac{2}{3}}\Lambda y^2\right) \label{equation_x_nonminimal}\\
\frac{dy}{dN}&=&-\frac{3y}{2}\left(\,(w-1)x^2+(w+1)(y^2-1)+\sqrt{\frac{2}{3}}\Lambda x\right)
\label{equation_y_nonminimal}\\
\frac{d\Lambda}{dN}&=&-\sqrt{6}(\Gamma-1)\Lambda^2 x, \hspace{\baselineskip}\text{where}
\hspace{\baselineskip}\Gamma = \frac{UU''}{{U'}^2} \label{equation_Lambda_nonminimal}.
\end{eqnarray}

If $\Lambda(\Phi)$ were invertible, we could express $\Gamma$ as a function of $\Lambda$ to close the system. In our case
\be \Lambda(\Phi)=\lambda\cosh(\sqrt{\xi}\Phi)+4\sqrt{\xi}\tanh(\sqrt{\xi}\Phi),\ee

which is not invertible. So we consider two {{limits; namely,}}  $\sqrt{\xi}|\Phi| \gg 1$ with $\Phi < 0$ and $\Phi > 0$. As mentioned earlier, the region $\sqrt{\xi}\Phi \ll 1$ corresponds to the usual exponential and we can show, using the variables (\ref{VARIABLES-1}), that it does not contain attractors or other interesting features. When $\sqrt{\xi}|\Phi| \gg 1$ with $\Phi < 0$
\be\label{Lambda_Gamma_nonminimal_<0}
\Lambda \simeq \frac{\lambda}{2}\exp(-\sqrt{\xi}\Phi)-4\sqrt{\xi}
\hspace{\baselineskip}\text{and}\hspace{\baselineskip}
\Gamma -1\simeq\frac{4\xi+\Lambda\sqrt{\xi}}{\Lambda^2}.\ee

\vfill\newpage\vfill

In particular, we have $\Lambda \gg -4\sqrt{\xi}$ in this limit, so that the approximation breaks when $\Lambda \sim -4\sqrt{\xi}$. With this expression of $\Gamma$, the system (\ref{equation_x_nonminimal})-(\ref{equation_Lambda_nonminimal}) has the following fixed points
\begin{table}[ht]
\begin{center}\begin{tabular}{l|c|c|c|r}
& $x_*$ & $y_*$ & $\Lambda_*$ & eigenvalues\\
\hline\hline
$O$ &$0$ & $0$ & $-$ & $0, -3/2(1-w), 3/2(1+w)$\\
\hline
$A$ & $0$ & $1$ & $0$ & $-3(1+w)$\\
&&&&$ -3/2\pm1/2\sqrt{3(3-16\xi)}$\\
\hline\hline
$B_\pm$ & $\pm 1$ & $0$ & $\Lambda$ & $3(1-w), 3\mp\sqrt{3/2}\Lambda$\\
\hline
$C_\Lambda$ & $1/\Lambda\sqrt{3/2}(1+w)$ & $1/\Lambda\sqrt{3/2(1-w^2)}$ & $\Lambda$ & $-3/4(1-w)\pm \kappa$\\
\hline
$A_\Lambda$ & $\Lambda/\sqrt{6}$ & $\sqrt{1-\Lambda/6}$ & $\Lambda$ & $-3(1+w) + \Lambda^2, -3+\Lambda^2/2$\\
\end{tabular}\end{center}
\caption{Fixed points of the system (\ref{equation_x_nonminimal})-(\ref{equation_Lambda_nonminimal}). The quantity $\kappa$ is given in (\ref{kappa}). In the column for $\Lambda_*$, we put a $-$ when $\Lambda$ is unconstrained by the equations, and a $\Lambda$ when we ignored equation (\ref{equation_Lambda_nonminimal}).}
\label{fixed_points_nonminimal}
\end{table}

In this table, $\kappa$ is given by:
\be \label{kappa}
\kappa=\frac{3}{4\Lambda}\sqrt{(1-w)\left(24(1+w)^2-(7+9w)\Lambda^2\right)}.\ee

Also, in the three last lines, we gave the fixed points of the reduced system (\ref{equation_x_nonminimal})-(\ref{equation_y_nonminimal}), treating $\Lambda$ as a constant. Of course, in this case, we recover the fixed points of the usual quintessence in an exponential potential. This will allow us to study transient regimes and the behaviour when $\Lambda \rightarrow \pm \infty$. Let us comment on some of the fixed points of Table \ref{fixed_points_nonminimal}.
\begin{itemize}
\item[$\bullet$]The only stable point is $A$. It corresponds to the minimum of the potential, because $\Lambda = 0 \leftrightarrow U' = 0$ from (\ref{VARIABLES-2}). At this point, we have $\Omega_\phi=1$ and  accelerated expansion.
\item[$\bullet$]The point ${B}_-$ is always unstable. The point ${B}_+$ is only unstable if $\sqrt{3/2}\Lambda<3(1-w)$. These points are past attractors corresponding to $\Omega_\phi = 1$ and $w_\phi = 1$. Numerically, running time backwards, the system is generically attracted to ${B}_+$ first. So, following (\ref{equation_Lambda_nonminimal}), $\Lambda$ increases. When $\sqrt{3/2}\Lambda\sim3(1-w)$, the point ${B}_+$ becomes a saddle and the system starts being attracted to $B_-$. There, $\Lambda$ decreases until $\Lambda \gg -4\sqrt{\xi}$ breaks.
\item[$\bullet$]The point $C_\Lambda$ is stable when $\Lambda^2 > 3(1+w)$. If $w=0$, it corresponds to $\Omega_\phi=3/\Lambda^2$ and, if $\Lambda$ is large, $\Omega_\phi \sim 0$. At this point, $\Lambda$ decreases (\ref{equation_Lambda_nonminimal}), so after some time it becomes unstable. Thus, $C_\Lambda$ is a transient point that can correspond to matter dominated era. 
\item[$\bullet$]The point $A_\Lambda$ is stable when $\Lambda^2 < 3(1+w)$. This is the point towards which the system is attracted after $C_\Lambda$. At this point, $\Lambda$ keeps on decreasing and $A_\Lambda$ converges to $A$.
\end{itemize}

\vfill\newpage\vfill

When $\sqrt{\xi}|\Phi| \gg 1$ with $\Phi > 0$, some signs are changed in equation (\ref{Lambda_Gamma_nonminimal_<0}); namely,
\be \label{Lambda_Gamma_nonminimal_>0}
\Lambda \simeq \frac{\lambda}{2}\exp(\sqrt{\xi}\Phi)+4\sqrt{\xi}
\hspace{\baselineskip}\text{and}\hspace{\baselineskip}
\Gamma -1\simeq\frac{4\xi-\Lambda\sqrt{\xi}}{\Lambda^2}.\ee

The fixed points $B_\pm, C_\Lambda, A_\lambda$ of the reduced system (\ref{equation_x_nonminimal})-(\ref{equation_y_nonminimal}) are the same as in Table \ref{fixed_points_nonminimal} in this case, but the behavior is a little different. In particular, the point $C_\Lambda$ is still stable when $\Lambda^2 > 3(1+w)$, but because of the sign change in front of $\Lambda$ in $\Gamma-1$ (\ref{Lambda_Gamma_nonminimal_>0}), now $\Lambda$ increases. So the point $C_\Lambda$ remains stable and is the future attractor of the system. This attractor has $\Phi$ increasing indefinitely. It corresponds to the field rolling down the "double exponential" tail of the potential (\ref{EFPOT}). Since $\Omega_\phi \rightarrow 0$, we do not obtain accelerated expansion in this regime.

\vfill

With this analysis, we now know that the trajectories that lead to a realistic background evolution, with matter domination followed by accelerated expansion, are with the field starting around the minimum of (\ref{EFPOT}). Note that this minimum is only present when $4\xi > \lambda^2$. We can use the knowledge that matter domination corresponds to $C_\Lambda$ on Table \ref{fixed_points_nonminimal} to compute the initial position of the field that leads to a given number of e-folds before accelerated expansion. There are two such positions $\Phi_i^L$ (left) and $\Phi_i^R$ (right), one on each side of the minimum. In the case of $\Phi_i^L$, we can say that $\Lambda$ is large initially to approximate $\Gamma -1 \simeq \sqrt{\xi}/\Lambda$ in (\ref{Lambda_Gamma_nonminimal_<0}). With this and using the value of $x_*$ corresponding to $C_\Lambda$ in Table \ref{fixed_points_nonminimal}, we obtain from (\ref{equation_Lambda_nonminimal}):
\begin{equation}
\frac{d\Lambda}{dN}\simeq -3(1+w)\sqrt{\xi} \hspace{\baselineskip}\rightarrow\hspace{\baselineskip}
\Lambda \simeq -3(1+w)\sqrt{\xi}N + \Lambda_i\,.
\end{equation}

Then, since the minimum corresponds to $\Lambda = 0$, we reach it after $N_f$ e-folds if initially
\begin{equation}\label{phi_initial_L_nonminimal}
\Phi_i^L \simeq	-\frac{1}{\sqrt{\xi}}\log\left(6(1+w)\frac{\sqrt{\xi}}{\lambda}N_f\right).
\end{equation}

For $\Phi_0^R$, we can approximate $\Lambda \simeq -4\sqrt{\xi}$ {and} write $\Lambda = -4\sqrt{\xi} + \Lambda'$, with small $\Lambda'$, to get
\begin{equation}
\frac{d\Lambda'}{dN} \simeq \frac{3}{4}(1+w)\Lambda' 
\hspace{\baselineskip}\rightarrow\hspace{\baselineskip}
\Lambda \simeq -4\sqrt{\xi}+(\Lambda_i+4\sqrt{\xi})\exp\left(\frac{3}{4}(w+1)N\right).
\end{equation}

Once again, we reach the minimum $\Lambda = 0$ after $N_f$ e-folds provided that initially
\begin{equation}\label{phi_initial_R_nonminimal}
\Phi_i^R \simeq -\frac{1}{\sqrt{\xi}}\left(\ln\left(\frac{8\sqrt{\xi}}{\lambda}\right)-\frac{3}{4}(1+w)N_f\right).
\end{equation}

These expressions work reasonably well when compared to numerical computations.

\vfill

\begin{figure}[ht]\begin{center}
\includegraphics[scale=.8]{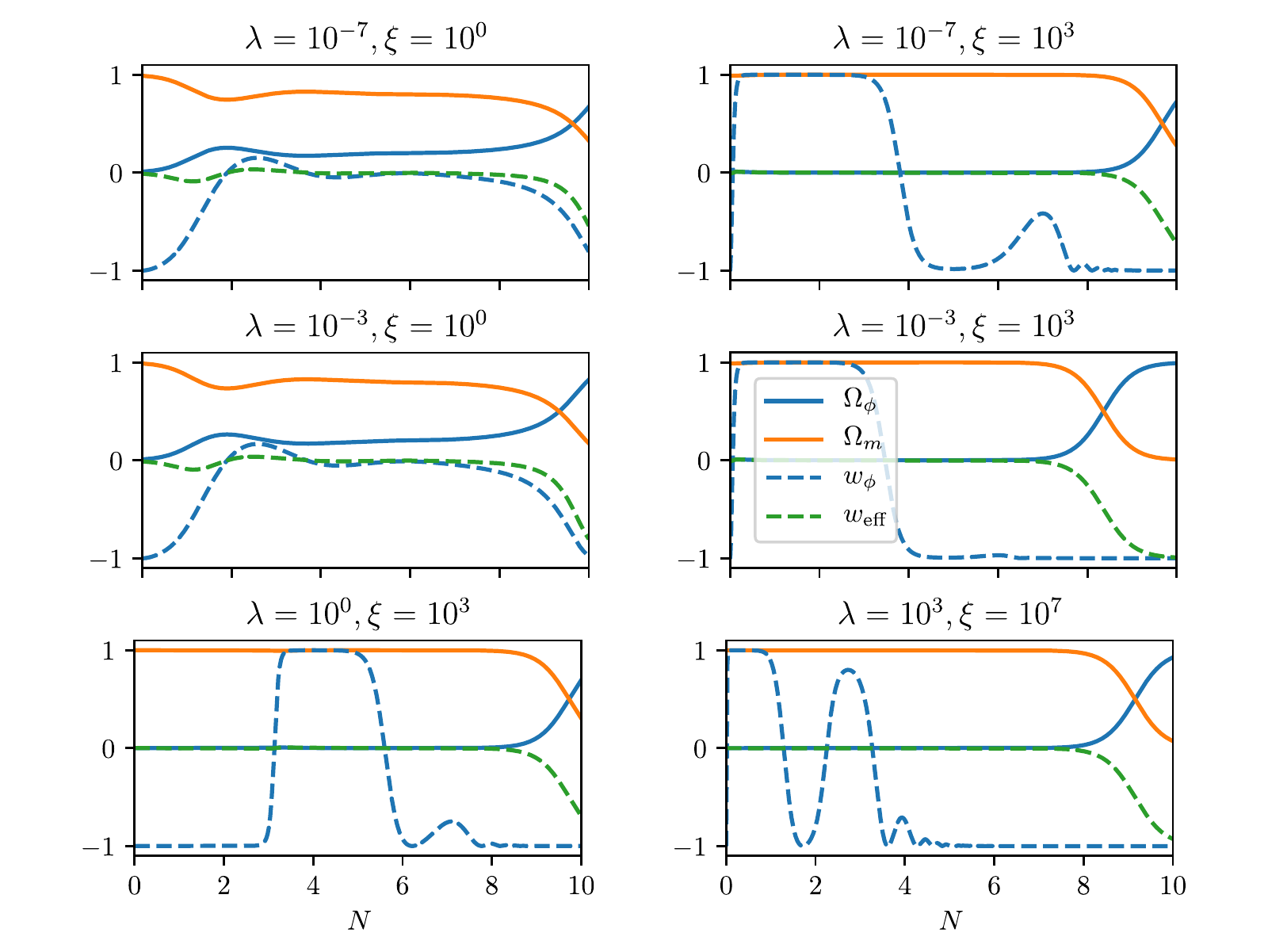}
\end{center}
\caption{Evolution of the scalar field parameters (\ref{scalar_parameters_nonminimal}) following from the system (\ref{equation_x_nonminimal})-(\ref{equation_Lambda_nonminimal}), for $w = 0$ and different values of $\lambda, \xi$. We denoted $\Omega_m = 1-\Omega_\phi$. We choose initial $x_i, y_i$ to have $\Omega_{ \phi}<0.2$ initially and $\Lambda_i$, or equivalently $\Phi_i$ to obtain between $8$ and $10$ e-folds of matter domination.}
\label{evolution_nonminimal}
\end{figure}

In Figure \ref{evolution_nonminimal}, we show the evolution of the scalar field parameters (\ref{scalar_parameters_nonminimal}) following from the system (\ref{equation_x_nonminimal})-(\ref{equation_Lambda_nonminimal}) for different values of $\lambda, \xi$, and initial $\Phi_i^R$ given by (\ref{phi_initial_R_nonminimal}) (we obtain similar results using $\Phi_i^L$ of (\ref{phi_initial_L_nonminimal}), except for the two first plots in the left as explained below). 

\vfill\newpage\vfill

The evolutions we observe in Figure \ref{evolution_nonminimal} confirm the previous analysis. The system starts being attracted to point $C_\Lambda$ of Table \ref{fixed_points_nonminimal}. At this point, $\Omega_\phi = 3/\Lambda^2$, so in particular if the evolution starts with the field at $\Phi_i^R$ of (\ref{phi_initial_R_nonminimal}), as it does in Figure \ref{evolution_nonminimal}, we have $\Lambda\simeq -4\sqrt{\xi}$ and $\Omega_\phi\simeq3/(16\xi)$ during matter domination. If $\xi$ is not large enough (for instance $\xi = 1$ as in two of the plots), we have $\Omega_\phi$ quite far from $0$. If the evolution starts with the field at $\Phi_i^L$ of (\ref{phi_initial_L_nonminimal}), we have $\Lambda \gg 1$ during matter evolution and $\Omega_\phi \simeq 0$ irrespectively of the value of $\xi$. In any case, the system ends up attracted to point $A$ of Table \ref{fixed_points_nonminimal}, corresponding to accelerated expansion.

\vfill

{In Table \ref{numerical_results_nonminimal}, we give some values of the parameters associated to Figure \ref{evolution_nonminimal}. These results confirm the analysis we made so far. First, the value of $M$ tuned numerically to have the observed Hubble rate today is indeed linked to $\lambda, \xi$ by (\ref{lambda_M_xi_nonminimal}), namely $\lambda^2M^2/\xi \sim 10^{-23}\ \mathrm{GeV}^2$. Then, the values $\Phi_0^{L, R}$ tuned to have $\sim10$ e-folds of matter domination approximately correspond to the ones computed in (\ref{phi_initial_L_nonminimal}) and (\ref{phi_initial_R_nonminimal}), and they do not depend much on the initial scalar energy density. Therefore, for each value of $\lambda, \xi$, the initial value of the field is pretty much entirely determined by the duration requirement of the matter dominated era.}

\vfill\newpage\vfill

\begin{table}[ht]
\begin{center}\begin{tabular}{c|c|c||c|c|c||c}
$\lambda$ & $\xi$ & $M\ (\mathrm{GeV})$ & $\Omega_0^{1/2}$ & $\Phi_0^L$ & $\Phi_0^R$ & $\dot\Phi_{\mathrm{max}}^2\ (\mathrm{GeV^4})$\\
\hline\hline &&&&&&\\
$10^{-7}$ & $10^{0} $ & $4\times 10^{-5}$ & $10^{-1}$ & $-20.4$ & $-11$ & $10^{-38}$\\
$$ & $ $ & $$ & $10^{-3}$ & $-20$ & $-14$ & $10^{-43}$\\
$$ & $10^{3} $ & $1\times 10^{-3}$ & $10^{-1}$ & $-0.75$ & $-0.46$ & $10^{-39}$ \\
$$ & $ $ & $$ & $10^{-3}$ & $-0.74$ & $-0.53$ & $10^{-45}$ \\
$$ & $10^{7} $ & $1\times 10^{-1}$ & $10^{-3}$ & $-0.0089$ & $-0.007$ & $10^{-47}$ \\
&&&&&&\\\hline &&&&&&\\
$10^{-3}$ & $10^{0} $ & $1\times 10^{-8}$ & $10^{-1}$ & $-11.2$ & $-2$ & $10^{-38}$ \\
$$ & $ $ & $$ & $10^{-3}$ & $-10.9$ & $-4$ & $10^{-42}$ \\
$$ & $10^{3} $ & $3\times 10^{-7}$ & $10^{-1}$ & $-0.46$ & $-0.2$ & $10^{-40}$ \\
$$ & $ $ & $$ & $10^{-3}$ & $-0.45$ & $-0.24$ & $10^{-47}$ \\
$$ & $10^{7} $ & $3\times 10^{-5}$ & $10^{-3}$ & $-0.006$ & $-0.004$ & $10^{-47}$ \\
&&&&&&\\\hline &&&&&&\\
$10^0$ & $10^{3} $ & $3\times 10^{-10}$ & $10^{-3}$ & $-0.23$ & $-0.0004$ & $10^{-45}$ \\
$$ & $10^{7} $ & $3\times 10^{-8}$ & $10^{-3}$ & $-0.0038$ & $-0.0017$ & $10^{-47}$ \\
&&&&&&\\\hline &&&&&&\\
$10^{3}$ & $10^{7} $ & $3\times 10^{-11}$ & $10^{-3}$ & $-0.0016$ & $-$ & $10^{-47}$ \\
\end{tabular}\end{center}
\caption{{Value of the parameters associated to some of the evolutions of Figure \ref{evolution_nonminimal}. The value of $M$ is tuned to obtain the current value of the Hubble rate at matter-scalar equality. The initial scalar energy density $\Omega_0$ is varied between $10^{-2}$ and $10^{-6}$. The values $\Phi_0^{L, R}$ are the two initial positions of the field tuned to obtain between $8$ and $10$ e-folds of matter domination. The value $\dot\Phi_{\mathrm{max}}^2$ is the maximum value that $\dot\Phi^2$ attains during the evolution and will be used to discuss the importance of the $\alpha(\nabla\Phi)^4$ term.}}
\label{numerical_results_nonminimal}
\end{table}

\vfill

{In the last line of Table \ref{numerical_results_nonminimal}, there is no value for $\Phi_0^R$. This is because this point, there is not enough distance between the local minimum and the local maximum of the potential $U$ (\ref{EXTREMA}) for the field to roll during the $8-10$ e-folds of matter domination. So above $\lambda\sim10^3$, the field can only start from the left of the local minimum to obtain the desired evolution.}

\vfill

For the usual quintessence in an exponential potential, we have $w_\phi = \lambda^2/3 - 1$ during late time scalar domination, so the experimental observation that $-1 < w_\phi < -0.95$ implies that $\lambda < 0.34$. Also, $M$ is of the order of $H_0^{1/4}$, that is $M\sim 10^{-11}\ \mathrm{GeV}$ in this model. In our case, $\lambda$ can take any value as long as $4\xi > \lambda^2$, and $M$ is given by $M \sim \sqrt{\xi}/\lambda \times 10^{-12}\ \mathrm{GeV}$, so it can be brought to much higher values, for instance up to $10^{-1}\ \mathrm{GeV}$ for $\lambda=10^{-7}$ and $\xi = 10^7$.

\vfill\newpage\vfill

Summarizing, we {saw that in the presence of the non-minimal coupling, the system can start in the vicinity of a transient point corresponding to matter domination (for initial values of the scalar field close to the local minimum, either on one side or the other) and evolve towards the late time attractor corresponding to accelerated expansion. This can be realized for a wide range of the $\xi$-parameter values. Also, it is clear in Table \ref{numerical_results_nonminimal} that the scalar field values can stay subplanckian as long as $\xi \gtrsim 10$}.

\vfill

\paragraph{Initial conditions and inflation.} In equation \eqref{phi_initial_L_nonminimal} and \eqref{phi_initial_R_nonminimal}, we gave expressions for the initial position of the field that give a desired number of e-folds before accelerated expansion. It turns out that the former can be linked to inflation. Let us go back to the Jordan frame action \eqref{starting_point_action}, and suppose that during inflation, $R\simeq12H_\mathrm{inf}^2$ is a constant, determined by the inflationary dynamics and independent of $\phi$. In this case, the nonminimal coupling $\xi\phi^2R$ enters in the potential of $\phi$
\begin{equation}
\tilde V(\phi) = V(\phi) - 6\xi H_\mathrm{inf}^2\phi^2,
\end{equation}

and $\phi$ will roll down to the minimum of this new potential. As a consequence, the minimum of $\tilde V$ is a natural initial condition for the field $\phi$ to start rolling after inflation, at the beginning of the radiation dominated era. For our exponential potential $V(\phi) = M^4\exp(-\lambda\phi)$, this minimum is located at
\begin{equation}
\phi_{\mathrm{min}} = \frac{1}{\lambda}W_{-1}\left(-\frac{\lambda^2\kappa^2M^4}{12\xi H_\mathrm{inf}^2}\right),
\end{equation}

where $W_{-1}$ is the lower branch of the Lambert function. Now, remind that $\lambda^2M^2/\xi \sim 10^{-23}\ \mathrm{GeV}^2$ to fit the observed value of $H_0$ \eqref{lambda_M_xi_nonminimal}, so we can write
\begin{equation}
\phi_{\mathrm{min}} = \frac{1}{\lambda}W_{-1}\left(-\left(\frac{M}{H_\mathrm{inf}}\right)^2\times10^{-60}\right).
\end{equation}

\vfill

For reasonable values of $M$ obtained in Table \ref{numerical_results_nonminimal} and reasonable values $H_\mathrm{inf}\sim10^4-10^{13}\ \mathrm{GeV}$, the argument of $W_{-1}$ is small, so we can approximate $W_{-1}(-x)\simeq \log(x)$. Finally, the initial position \eqref{phi_initial_L_nonminimal} was given in terms of the canonical field $\Phi$, which was defined as a function of $\phi$ in \eqref{FI}. Using this definition and the approximation $\sinh^{-1}(x)\simeq-\log(-2x)$ for large negative $x$, we get
\begin{equation}
\Phi_{\mathrm{min}} \simeq -\frac{1}{\sqrt{\xi}}\log\left(\frac{2\sqrt\xi}{\lambda}\log\left(\left(\frac{H_\mathrm{inf}}{M}\right)^2\times10^{60}\right)\right).
\end{equation}

This is to be compared with \eqref{phi_initial_L_nonminimal}. First of all, it has the same dependence in $\lambda$ and $\xi$. Now, in the case where the field rolls during $N_r$ e-folds of radiation dominated era and $N_m$ e-folds of matter dominated era, we just have to replace $(1+w)N_f$ by $4/3N_r + N_m$ in \eqref{phi_initial_L_nonminimal}. Therefore, if the field starts in this inflationary-driven minimum at the beginning of the radiation dominated era, we will have the following link between $N_r, N_m$ and $H_\mathrm{inf}, M$
\begin{equation}
4N_r + 3N_m = 138+2\log\left(\frac{H_\mathrm{inf}}{M}\right).
\end{equation}

\vfill\newpage\vfill

If we take $H_\mathrm{inf} \sim 10^4 - 10^{13}\ \mathrm{GeV}$ and $M\sim 10^{-11} - 10^{-1}\ \mathrm{GeV}$, and insist that $N_m\sim 8-10$, we obtain $N_r\simeq33-56$. If we fix $H_\mathrm{inf}$ to the "natural" value $10^{13}\ \mathrm{GeV}$, we get $N_r\simeq43-56$. Those are very reasonable values for the number of e-folds of radiation domination.

\vfill

\section{Combination of both}

\vfill

In this section we shall consider the combined presence of both corrections and check whether the dynamics generated by the non-minimal coupling $\xi\phi^2R$ are modified by the quadratic correction $\alpha R^2$, provided {that} the $\alpha$ parameter takes reasonable values. There are two ways that $\alpha$ enters in the action (\ref{action_alpha_nonminimal}); namely, through $\alpha U$ and through $\alpha(\nabla\Phi)^4$. Let us start with the latter. For this term to be non-negligible, it has to be larger than the kinetic term $(\nabla\Phi)^2$. This would happen if at any point $\alpha\dot\Phi^2 > 1$. In the numerical computations of Figure \ref{evolution_nonminimal}, with values of the parameter that reproduce the current $H_0$, we have checked that the largest $\dot\Phi^2$ gets is around $10^{-114}$ in Planck units, {see the last column of Table \ref{numerical_results_nonminimal}}. So we can be confident that the term $\alpha(\nabla\Phi)^4$ does not play a role on the late evolution for reasonable values of $\alpha$.

\vfill

The other way $\alpha$ can play a role is if $\alpha U > 1$, that is, in regions where the potential is large. According to the analysis that follows (\ref{EFPOT}), there are two such regions: the divergent part where $\Phi \rightarrow -\infty$ and the local maximum $U_{\mathrm{max}}$. According to (\ref{potential_U_limits}), we have $U_{\mathrm{max}}\simeq M^4$, and according to (\ref{lambda_M_xi_nonminimal}) we have $M^4 \sim \xi^2/\lambda^4\times10^{-122}$ in Planck units. So even with quite extreme values $\xi\sim10^{10}$ and $\lambda\sim10^{-10}$ we would still need $\alpha\sim10^{62}$ for it to play a role close to $U_{\mathrm{max}}$.

\vfill

So the only remaining possibility to have $\alpha U > 1$ and $\alpha$ playing a role is in the region $\Phi \rightarrow -\infty$ where $U(\Phi)$ diverges according to (\ref{EFPOT}). What is interesting is that it diverges very quickly, in an "exponential of exponential" fashion, so the field does not need to go far for $U$ to be really large, even if $M^4$ is small to match the current value $H_0$. This region of large negative $\Phi$ is not the one where the evolutions of Figure \ref{evolution_nonminimal} take place, but it is explored by the past attractor of (\ref{equation_x_nonminimal})-(\ref{equation_Lambda_nonminimal}), when the system follows the point $B_+$ of Table \ref{fixed_points_nonminimal}. {Moreover, if we link the initial position of the field with the end of inflation as in the end of Section \ref{sec:non-minimal_coupling}, it has to start further away in the negative to accomodate the e-folds of radiation domination, in a region where $U$ is definitely large.} As such, the terms in $\alpha U$ do not play a role during matter domination and after, but could have done so in the beginning of the radiation era.

\vfill

Summarizing, we conclude that in the combined presence of both corrections, although the quadratic term {provides a new past attractor} corresponding to an effective radiation phase with $w_{\phi}=1/3$, it does not play any role during matter domination. Thus, the system can start {near the transient point corresponding to matter domination and evolve, making a transition to the late time attractor corresponding to accelerated expansion, in a way that is driven exclusively by the non-minimal coupling correction.}

\vfill

\section{Conclusions}

\vspace{\baselineskip}

We have considered a scalar field $\phi$ coupled to gravity in the presence of corrections to the Einstein-Hilbert action, namely the quadratic $\alpha R^2$ term and a non-minimal coupling term $\xi\phi^2R$, in the framework of the Palatini formulation. We focused on the case of an exponential self-interaction potential for the scalar field $V(\phi)=M^4\exp({-\lambda \phi})$. We studied the dynamics of the system {in the goal of obtaining} a {\textit{``realistic"}} evolution of the Universe, with a long enough matter domination era followed by accelerated expansion. For this we made use of dynamical systems methods. 

\vspace{\baselineskip}

We showed that the quadratic correction cannot play any role in the late-time dynamics, unless the parameter $\alpha$ were to take absurdly large values; although it could play a role at early times. In contrast, the non-minimal coupling correction can drive the evolution of the system {from the vicinity of a transient point corresponding to matter domination towards a late time attractor corresponding to accelerated expansion. The non-minimal coupling induces a local minimum in the potential, provided that $4\xi>\lambda^2$. We found that the only viable trajectories are the ones where the canonical field starts away from the minimum, at a distance that is dictated by the number of e-folds of matter domination, and ends up at the minimum, thus acting as an effective cosmological constant. This minimum corresponds to accelerated expansion, while the field rolling towards the minimum corresponds to matter domination. It should be noted that during this evolution the scalar field can stay well within subplanckian values. {Also, linking the initial position of the scalar field with inflation, we found an interesting relation between the number of e-folds of radiation/matter domination and the energy scales of inflation/quintessence.}}

\vspace{\baselineskip}

\section*{Acknowledgments}

K. T. aknowledges usefull discussions with Alexandros Papageorgiou.

\vfill

\bibliographystyle{kp}
\bibliography{bibliographie}

\end{document}